\documentclass{aa}  

\usepackage{graphicx}
\usepackage{txfonts}
\usepackage{hyperref}
\usepackage{xcolor}
\usepackage[normalem]{ulem}

\urldef{\stixads}\url{https://ui.adsabs.harvard.edu/abs/2020A%26A...642A..15K/abstract}
\urldef{\eisdownload}\url{http://solarb.mssl.ucl.ac.uk/SolarB/SearchArchive.jsp}
\urldef{\xrtdownload}\url{http://sdc.uio.no/search/form}
\urldef{\stixdownload}\url{https://pub023.cs.technik.fhnw.ch/view/list/bsd}
\urldef{\aiadownload}\url{http://jsoc.stanford.edu/}
\urldef{\STIXGSW}\url{https://github.com/i4Ds/STIX-GSW}

\begin{document}

   \title{Solar flare hard X-rays from the anchor points of an eruptive filament}

   \author{Muriel Zo\"e Stiefel \inst{1}
          \and
          Andrea Francesco Battaglia \inst{1}\fnmsep\inst{2}
          \and
          Krzysztof Barczynski \inst{1}\fnmsep\inst{3}
          \and
          Hannah Collier \inst{1}\fnmsep\inst{2}
          \and
          Anna Volpara\inst{4} 
          \and 
          Paolo Massa\inst{5}
          \and 
          Conrad Schwanitz \inst{1}\fnmsep\inst{3}
          \and
          Sofia Tynelius \inst{1}
          \and
          Louise Harra \inst{3}\fnmsep\inst{1}
          \and
          S\"am Krucker \inst{2}\fnmsep\inst{6}
          }

   \institute{ETH Z\"{u}rich,
              R\"{a}mistrasse 101, 8092 Z\"{u}rich Switzerland\\
              \email{muriel.stiefel@gmail.com} 
        \and
            University of Applied Sciences and Arts Northwestern Switzerland, Bahnhofstrasse 6, 5210 Windisch, Switzerland
         \and
             PMOD/WRC, Dorfstrasse 33, CH-7260 Davos Dorf, Switzerland
        \and
            MIDA, Dipartimento di Matematica,  Universit\`{a} degli Studi di Genova, Via   Dodecaneso 35, 16146 Genova, Italy
        \and
            Department of Physics \& Astronomy, Western Kentucky University, Bowling Green, KY 42101, USA
        \and
            Space Sciences Laboratory, University of California, 7 Gauss Way, 94720 Berkeley, USA
}
   \date{Received September 23, 2022; accepted November 18, 2022}

 
  \abstract
   {We present an analysis of a GOES M1.8 flare with excellent observational coverage in UV, extreme-UV (EUV), and X-ray, including observations from the Interface Region Imaging Spectrograph (IRIS), from the Solar Dynamics Observatory (SDO) with the Atmospheric Imaging Assembly (AIA), from the Hinode/EUV Imaging Spectrometer (EIS), from the Hinode/X-ray Telescope (XRT), and from Solar Orbiter with the Spectrometer/Telescope for Imaging X-rays (STIX). Hard X-ray emission is often observed at the footpoints of flare loops and is occasionally observed in the corona. In this flare, four nonthermal hard X-ray sources are seen.}
   {Our aim is to understand why we can observe four individual nonthermal sources in this flare and how we can characterize the physical properties of these four sources.}
   {We used the multiwavelength approach to analyze the flare and characterize the four sources. To do this, we combined imaging at different wavelengths and spectroscopic fitting in the EUV and X-ray range.}
   {The flare is eruptive with an associated coronal mass ejection, and it shows the classical flare picture of a heated flare loop seen in EUV and X-rays, and two nonthermal hard X-ray footpoints at the loop ends. In addition to the main flare sources, we observed two outer sources in the UV, EUV, and nonthermal X-ray range located away from the main flare loop to the east and west. The two outer sources are clearly correlated in time, and they are only seen during the first two minutes of the impulsive phase, which lasts a total of about four minutes.}
   {Based on the analysis, we determine that the outer sources are the anchor points of an erupting filament. The hard X-ray emission is interpreted as flare-accelerated electrons that are injected upward into the filament and then precipitate along the filament toward the chromosphere, producing Bremsstrahlung. While sources like this have been speculated to exist, this is the first report of their detection.}


   \maketitle
%

\section{Introduction}

   Solar flares are observable as a brightening in emission over the entire electromagnetic spectrum with a duration of minutes to hours (for reviews, we refer to \citet{Benz_2017} and \citet{Fletcher_2011}). To date, a complete understanding of solar flares is lacking. Nevertheless, there are several well-observed aspects of solar flares that are described in various flare models. A famous model is the CSHKP model \citep{Carmichael_1964,Sturrock_1966,Hirayama_1974,Kopp_Pneuman_1976}, which is the basis of the standard model for solar flares. For many observed flares, the standard model accurately describes the basic physical process. In the following, we briefly summarise some aspects of solar flares that are explained by this model.
   
   A solar flare is a release of magnetic energy. It is caused by magnetic reconnection in the solar corona. Several models suggest that magnetic reconnection is triggered by the eruption of a filament or a flux rope \citep{Hirayama_1974,Shibata_1995}. This filament forms part of the core of a coronal mass ejection (CME). It is often difficult to observe the magnetic flux ropes until they start to erupt. A part of the released flare energy goes into the acceleration of particles in the solar corona. These particles propagate along magnetic field lines, penetrating deeper into the atmosphere. As the atmosphere becomes denser when the solar surface is approached, there is enough Coulomb interaction between the accelerated electrons and the plasma of the atmosphere at the height of the chromosphere to produce significant Bremsstrahlung \citep[e.g., ][]{Kontar_2011}. In this phase of the flare, strong nonthermal emission in the hard X-ray (HXR) and microwave (MW) wavelength ranges is characteristically measured. The locations at which the electrons hit the chromosphere and emit Bremsstrahlung is commonly called the ribbons or footpoints of a flare loop. The measurement of HXR emission at the footpoints was first reported by \citet{Hoyng_1981}, and it is a well-established observational fact that the main HXR sources are from footpoints \citep[e.g., ][]{SaintHilaire_2008}. However, for dense flare loops, an additional nonthermal source is occasionally seen in the corona at the loop top \citep[e.g., ][]{Veronig_2004,Glesener_2020}. Since 2017, the Expanded Owens Valley Solar Array (EOVSA) is observing the Sun in the microwave (MW) range \citep[for new results, see, e.g., ][]{Chen_2020_2,Fleishman_2020,Gary_2018,Fleishman_2022}. New insights by \citet{Gary_2018} and \citet{Fleishman_2022} based on MW observations with a higher dynamic range than indirect-imaging HXR observations show that accelerated electrons are not only seen within the main flare loop, but also populate the surrounding field lines that are associated with the flare and the CME. The precipitating particles that reach the footpoints heat the chromosphere. The heated material moves upward into the flare loop \citep{Antonucci_1982}. This concept is known as chromospheric evaporation (sometimes more accurately called chromospheric ablation) and was first reported by \citet{Doschek_1980} and \citet{Feldman_1980}. The material inside the coronal loop becomes dense, leading to thermal Bremsstrahlung. The thermal emission can be measured, for example, in the soft X-ray range (SXR). In summary, HXR observations show  nonthermal emission at the footpoints of a flare loop, and the thermal emission comes from the flare loop itself \citep{Benz_2017}. In occulted flares, where the main flare footpoints are blocked by the solar limb, much fainter emission associated with the escaping CME is detected \citep{Krucker_2007,Glesener_2013,Lastufka_2019}. This indicates that nonthermal electrons during flares are not only injected toward the chromosphere and produce the flare ribbon HXR sources, but flare-accelerated electrons are also escaping into the core of the CME. Some of the upward injected electrons are therefore expected to precipitate to the anchor points of the erupting CME and produce HXRs away from the main flare ribbon, although the intensity is expected to be fainter than the HXR sources from the flare footpoints. It has indeed been suggested (e.g., \citet{Shibata_1995}) that there might be additional X-ray sources at the anchor points of the erupting filament, but no such HXR sources have been reported so far. This might be attributed to the low intensities of these sources in combination with the limited dynamic range of an indirect HXR imaging system, such as the one adopted by the Reuven Ramaty High Energy Solar Spectroscopic Imager \citep[RHESSI;][]{Lin_2002}. In a recent study of SOL2017-09-10, an X8.2-class limb flare, \citet{Chen_2020} for the first time reported nonthermal emission in the MW range at the anchor points of an erupting filament measured by EOVSA. RHESSI observed this event as well, but measured no HXR emission at the anchor points of the filament. Here we report the first detection of such sources in the HXR range. 
   
   In February 2020, the satellite Solar Orbiter \citep{Muller_2020} was launched with ten different instruments on board to observe the Sun in detail. One of the instruments is the Spectrometer/Telescope for Imaging X-rays \citep[STIX;][]{Krucker_2020}, which studies solar flares in the hard X-ray energy range of 4-150 keV. This enables STIX to observe the hottest thermal plasma with temperatures above $\sim$8 MK \citep{Battaglia_2021} and nonthermal electrons providing information about their energy distribution. STIX uses an indirect imaging technique similar to the one used for its predecessor instrument RHESSI \citep{Lin_2002}, which allows the measurement of a limited number of Fourier components (or visibilities) of the flaring X-ray source. STIX is essentially observing the Sun continuously since the start of 2021, providing us with a rich data set awaiting analysis. The basic idea of this work was to analyze solar flares by combining data of the new instrument STIX with the well-established instruments Hinode/EUV Imaging Spectrometer \citep[EIS;][]{Culhane_2007}, Hinode/X-ray Telescope \citep[XRT;][]{Golub_2007}, and Solar Dynamics Observatory/Atmospheric Imaging Assembly \citep[SDO/AIA;][]{Lemen_2012}. In this paper we present the analysis of the SOL2021-09-23UT15:28M1.8 flare, the largest jointly observed flare of the four different observatories, which was detected before October 2021. In this flare, we observed hard X-ray signatures from flare-accelerated electrons away from the main flare loop.  
   
   The paper is structured as follows. In Sect. \ref{Chap:Data processing} we present the various instruments that we used in the analysis and explain the basic data-processing steps. In Sect. \ref{Ch: Observations} we present the observations and analysis of the M1.8 GOES class flare. The analysis of the flare is motivated by the STIX measurements coordinated with Hinode observations. Furthermore, we carry out an imaging-spectroscopy analysis of this flare using STIX data. Finally, we discuss the observations in Sect. \ref{Ch: Discussion and conclusion} and draw conclusions. 

\section{Data processing} \label{Chap:Data processing}

   \begin{figure*}
   \centering
   \includegraphics[trim=90 90 50 90, clip,width=19cm]{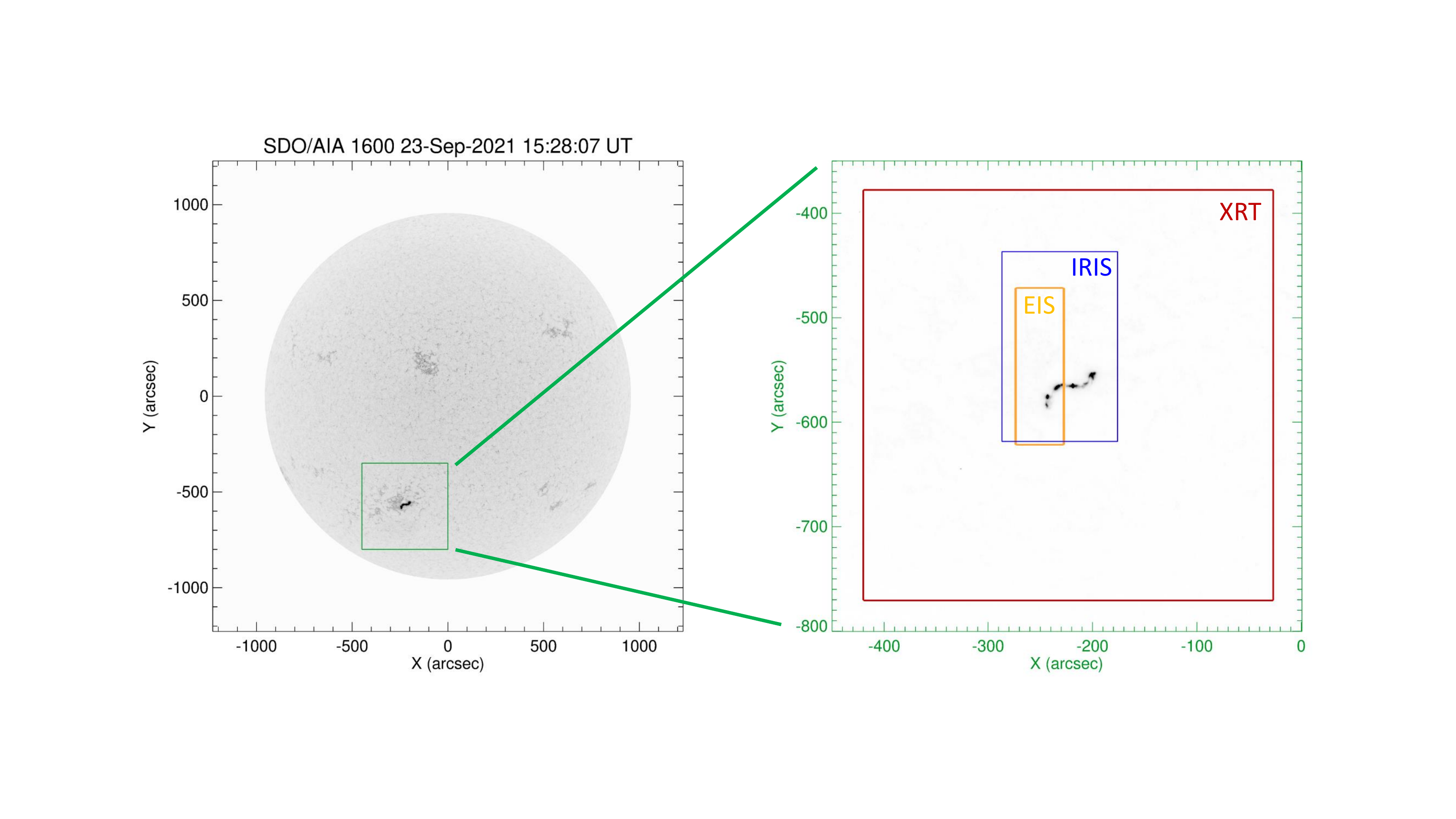}
   \caption{Flux measured by the AIA 1600 \r{A} filter. In the left panel the whole Sun is shown, and the right panel gives a close up around the flare location. The boxes in the right panel show the field of view of the instruments Hinode/EIS (yellow), Hinode/XRT (red), and IRIS (blue) for the SOL2021-09-23UT15:28 flare. STIX observes the whole solar disk from the viewing angle of Solar Orbiter (at an angle of about 33 degrees to the east).}
   \label{Fig: Field of view}%
    \end{figure*}
%

   \begin{figure*}
   \centering
   \includegraphics[trim=140 80 140 80, clip,width=0.8\textwidth]{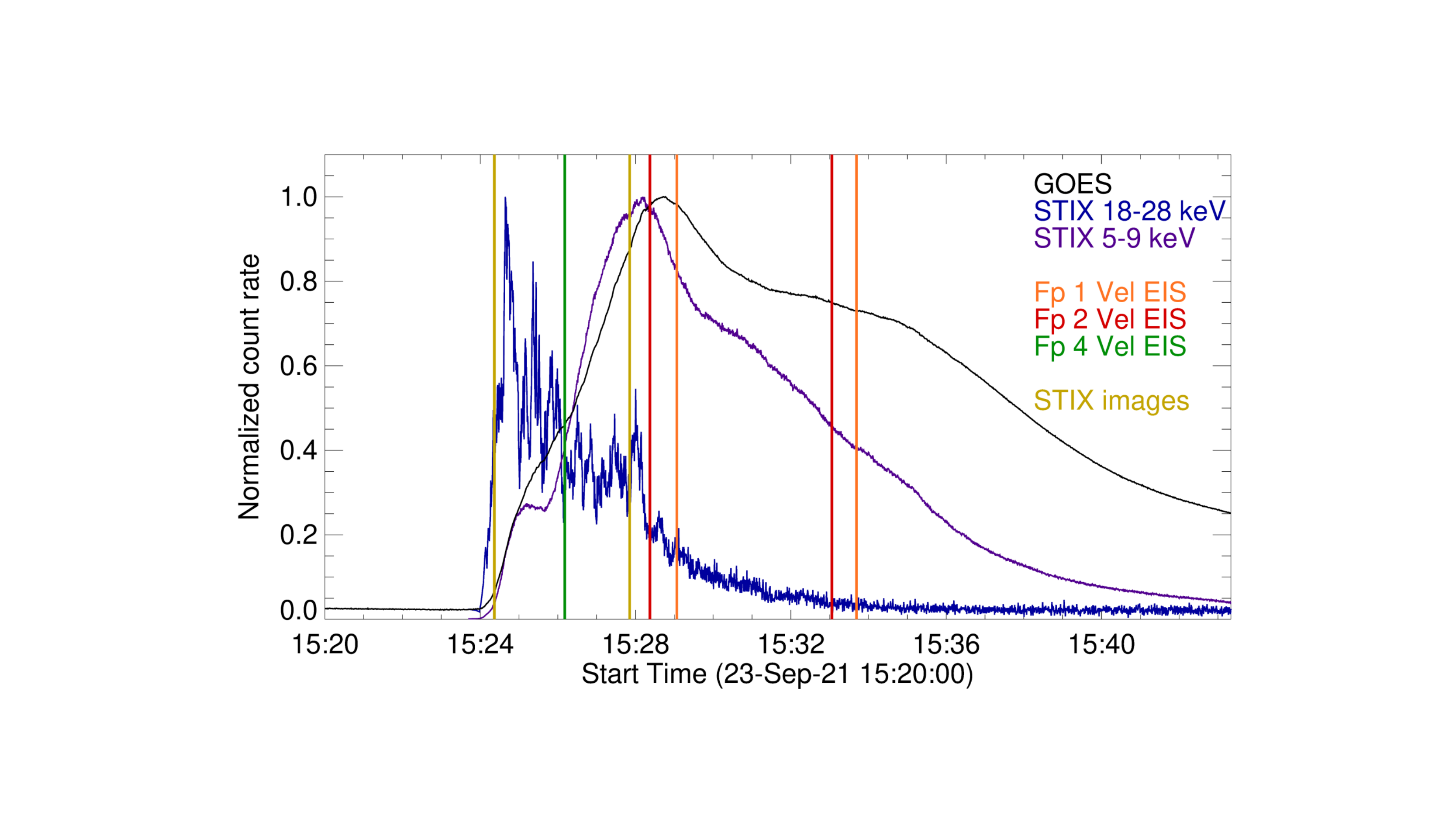}
   \caption{Intensity evolution of the flare measured by GOES (black), STIX thermal (violet), and STIX nonthermal (blue) over the whole Sun. The maximum peak of each curve is normalized to one. We added the time stamps with observational coverage for the velocity measurements; see Sect. \ref{Ch: Velocities}. The orange lines correspond to the time points when EIS measured at the position of Fp 1, the red lines when EIS measured at the position of Fp 2, and the green line when IRIS measured at the position of Fp 4. Furthermore, the integration times of the STIX images shown in Fig. \ref{Fig: non-thermal STIX image} and Fig. \ref{Fig: thermal geometric comparison} are delimited by the beige lines.}
   \label{Fig: Time evolution for Vel}%
    \end{figure*}

\subsection{Observation}
    We analyzed the M1.8 GOES class flare that occurred on 23 September 2021 at 15:28 UT. The instruments Hinode/XRT, Hinode/EIS, the Interface Region Imaging Spectrograph \citep[IRIS;][]{Pontieu_2014}, SDO/AIA, and Solar Orbiter/STIX observed the flare. Table \ref{Tab: Overview data} gives an overview of all the instruments and filters used for the analysis.
    
    The location of the flare on the solar disk is shown in Fig. \ref{Fig: Field of view} to the left. The image to the right shows a close-up of the flare site and the field of views of the instruments EIS, XRT, and IRIS. STIX observes the total solar disk from the viewing angle of Solar Orbiter. Figure \ref{Fig: Time evolution for Vel} shows an overview of the time evolution of the flare. In the following we discuss the calibration and data alignment that were carried out for each instrument before the analysis. 
    
   \begin{table}
      \caption[]{Summary of the different instruments and filters used for our analysis.}
         \label{Tab: Overview data}
         $$
         \begin{array}{lllll}
            \hline\hline
            \noalign{\smallskip}
            \mathrm{Instr.}  & \mathrm{Line/Filter} & \lambda & log T & \mathrm{Usage} \\
             & & \mathrm{[\r{A}]} & {[\mathrm{K}]} & \mathrm{[Section\;Number]} \\
            \noalign{\smallskip}
            \hline
            \noalign{\smallskip}
            \mathrm{EIS} & \mathrm{\ion{Fe}{xii}} & 195.12& 6.1 & 3.5\\
            \mathrm{EIS} & \mathrm{\ion{Fe}{xiii}} & 202.04 & 6.2 & 3.5\\
            \mathrm{EIS} & \mathrm{\ion{Fe}{xxiii}} & 263.76 & 7.1 & 3.2 \\
            \mathrm{EIS} & \mathrm{\ion{Fe}{xxiv}} & 192.15 & 7.2 & 3.2\\
            \noalign{\smallskip}
            \hline
            \noalign{\smallskip}
            \mathrm{XRT} & \mathrm{Be\; thick} & & \sim7 & 3.2\\
            \noalign{\smallskip}
            \hline
            \noalign{\smallskip}
            \mathrm{AIA} & \mathrm{AIA\; 131} & 131 & 5.6 & 2.7\\
            \mathrm{AIA} & \mathrm{AIA\; 171} & 171 & 6.8 & 3.4\\
            \mathrm{AIA} & \mathrm{AIA\; 193} & 193 & 6.1 & 3.3\;\&\;3.4\\
            \mathrm{AIA} & \mathrm{AIA\; 211} & 211 & 6.3 & 2.7\\
            \mathrm{AIA} & \mathrm{AIA\; 304} & 304 & 4.7 & 3.4\\
            \mathrm{AIA} & \mathrm{AIA\; 1600} & 1600 & 4 & 2.7\;\&\;3.2\;\&\;3.3\\
            \noalign{\smallskip}
            \hline
            \noalign{\smallskip}
            \mathrm{STIX} & \mathrm{4-150\; keV} & & \geq 6.9 & 3.1\;\&\;3.2\;\&\;3.6\\
            \noalign{\smallskip}
            \hline
            \noalign{\smallskip}
            \mathrm{IRIS} & \mathrm{C\; II} & 1335 & 4.6 & 3.5\\
            \mathrm{IRIS} & \mathrm{Si\; IV} & 1400 & 4.9 & 3.5\\
            \mathrm{IRIS} & \mathrm{Mg\; II} & 2796 & 3.9 & 3.5\\
            \noalign{\smallskip}
            \hline
         \end{array}
     $$ 
   \end{table}

\subsection{Hinode EIS}

    The instrument EIS \citep{Culhane_2007} is mounted on the satellite Hinode \citep{Kosugi_2007}. EIS observes the upper transition region and the solar corona in the two EUV wavelength bands 170-210 \r{A} and 250-290 \r{A}. The measured emission lines cover temperatures ranging from 0.04 MK to 20 MK.
        
    We downloaded the level-0 data from the Hinode/EIS website \footnote{\eisdownload}. For the analysis, we used the emission lines \ion{Fe}{xii} (195.12 \r{A}), \ion{Fe}{xiii} (202.04 \r{A}), \ion{Fe}{xxiii} (263.76 \r{A}), and \ion{Fe}{xxiv} (192.15 \r{A}). The EIS data calibration and data analysis was carried out using the standard procedures of the SSWIDL software. To calibrate the data, we used the routine eis\_prep. Additionally, we corrected the data for the orbital variation with the procedure eis\_wave\_corr. We determined the peak intensity and Doppler velocity maps from the EIS data set. To do this, we fitted a single Gaussian curve to the spectra using the procedure eis\_auto\_fit. For the emission lines \ion{Fe}{xiii}, \ion{Fe}{xxiii,} and \ion{Fe}{xxiv,} we additionally provided a template that restricted the wavelength range in which the Gaussian fit was to be performed using the procedure eis\_fit\_template. Using the fit parameters, we extracted the desired maps described above. 
        
    Because the emission line \ion{Fe}{xxiv} is known to be blended by other emission lines \citep{Zanna_2008}, we used \ion{Fe}{xxiii} as a reference map for the emission of \ion{Fe}{xxiv}. Based on the reference map, we set a minimum intensity that must be measured by the \ion{Fe}{xxiv} filter. All values below we set to zero, as we assume that this emission does not come from the \ion{Fe}{xxiv} emission line. 
    
    Because EIS has two different observation wavelength bands, we needed to align the maps of the different emission lines. To do this, we used the standard procedure eis\_ccd\_offset to find the offset between the maps. We used \ion{Fe}{xii} as a reference map and aligned all other lines to this map. 

\subsection{Hinode XRT}
    
    The instrument XRT \citep[][]{Golub_2007} is mounted on the satellite Hinode. XRT observes the solar corona in the soft X-ray energy range covering temperatures of several million kelvin. 
    
    We downloaded the level-0 data from the Hinode SDC Europe Archive website \footnote{\xrtdownload}. Then we calibrated the data and removed some of the CCD bias by using the procedure xrt\_prep. For the analysis, the data of the Be thick filter was used.

\subsection{Solar Orbiter STIX}
    
    The STIX level 1A data of the flare and the background file were downloaded from the STIX website \footnote{\stixdownload}. The complex visibility values were computed by means of the August 2022 version of the STIX Ground Software \footnote{\STIXGSW}. For the reconstruction of the STIX images, we used the Clean method \citep[stx\_vis\_clean;][]{Hogbom_1974}, the maximum entropy method \citep[MEM\_GE;][]{Massa_2020}, and the new procedure for forward fitting \citep[stx\_vis\_fwdfit\_pso;][]{Volpara_2022}. The rationale behind the forward-fitting method is the following. First, number and type (e.g., circular Gaussian, elliptical Gaussian, or loop) of the parametric shapes to be used to fit the visibilities were selected. Then, the values of the parameters that best fit the data (i.e., minimize the $\chi^2$) were determined by means of the particle swarm optimization algorithm \citep[PSO;][]{kennedy1995particle}. Finally, estimates of the uncertainty on the parameter values were provided by repeating the forward-fitting procedure 20 times, each time from visibilities perturbed with Gaussian noise, and by computing the standard deviation of the 20 results \citep{Volpara_2022}. Hence, the forward-fitting procedure returned not only a reconstructed image, but also the values of position, size, and flux of each source and the associated error. We refer to \citet{pianabook} and \citet{Massa_2022} for further details of the imaging methods that we used for the analysis presented in this paper.

\subsection{SDO AIA}
    
    The instrument AIA \citep[][]{Lemen_2012} is mounted on the SDO. AIA observes the solar atmosphere from the temperature minimum, via the chromosphere and the transition region, to the solar corona in seven EUV wavelength bands and two UV bands, and in the visible range. 
    
    We downloaded preprocessed data from the website of the Joint Science Operations Center (JSOC) \footnote{\aiadownload}. We used the keyword aia\_scale, which calibrates the data to level 1.5 before downloading. We downloaded the data of the UV wavelength band 1600 \r{A} and five EUV bands 193, 211, 171, 131, and 304 \r{A} in the time range 15:15 to 16:00 UT. 

\subsection{IRIS}
    
    The instrument IRIS \citep[][]{Pontieu_2014} is a multichannel imaging spectrograph observing the Sun in two far-ultraviolet (FUV) passbands, 1332-1358 \r{A} and 1389-1407 \r{A}, and in one near-ultraviolet (NUV) passband, 2783-2835 \r{A}. The passbands cover observations from the photosphere up to the lower corona. 
    
    We used the raster scans of IRIS from the wavelengths 1335 \r{A} (\ion{C}{ii}), 1400 \r{A} (\ion{Si}{iv}), and 2796 \r{A} (\ion{Mg}{ii}). In order to determine the peak intensity and Doppler velocity maps, we used a single Gaussian curve to fit the emission lines. 

\subsection{Data alignment}
    
    For the alignment of all data sets, we used the AIA maps as a reference. For the alignment of EIS \ion{Fe}{xii} and AIA 193 \r{A}, we used the procedure eis\_aia\_offset as a first estimate and improved the alignment by using cross correlation. We used the IDL procedure c\_correlate to calculate the cross correlation between the EIS and AIA map. By slightly shifting the EIS map with respect to the AIA map, we searched for the best-matching shift based on the values of the cross correlation. We aligned XRT with AIA 131 \r{A} and IRIS with AIA 1600 \r{A}. For both we used cross correlation for the alignment, as described above for EIS. The alignment is not straightforward, and therefore the accuracy might be off by a few arcseconds.
    
    For the alignment of STIX with AIA, we first had to correct for the different viewing angles between STIX and AIA. The separation angle between STIX and SDO was about 33 degrees on that day. To correct for this, we reprojected the AIA 1600 \r{A} frame to the viewing angle of STIX. Then, we corrected the location of the STIX reconstructions based on an estimate of the instrument pointing provided by the aspect system \citep{warmuth2020stix}. No further shifts to the image locations were applied.
    
   \begin{figure*}[htbp!]
   \centering
   \includegraphics[trim=40 110 40 110, clip,width=18cm]{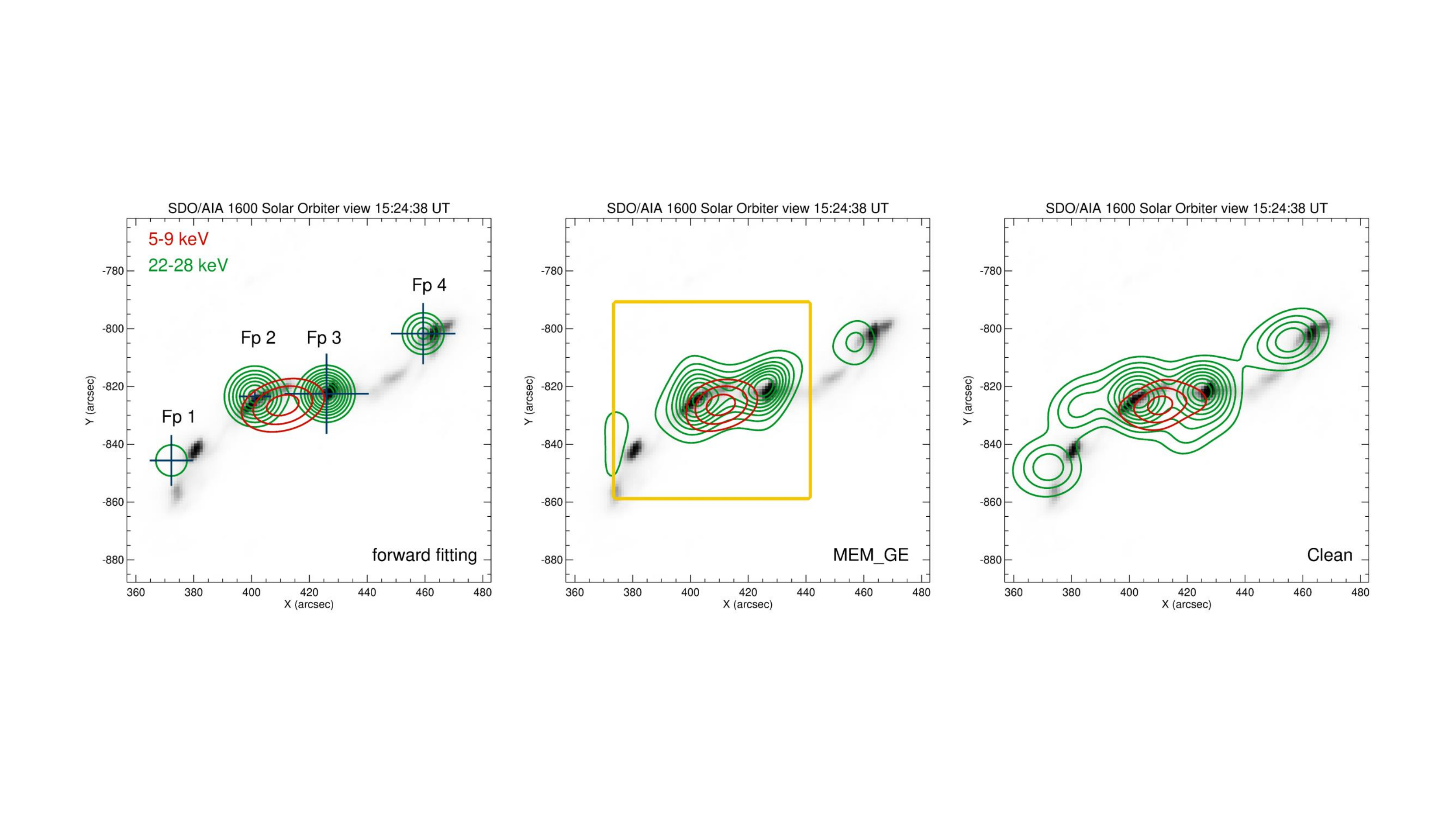}
   \caption{Level curves of the STIX reconstructions obtained with forward fit, MEM\_GE, and Clean (\emph{left}, \emph{middle}, and \emph{right} panels, respectively), overlaid on the reprojected AIA 1600 \r{A} map. The green contours (20-90\% of the peak value) represent the nonthermal emission reconstructed in the 22-28 keV range, and the red contours (50, 70, and 90\% of the peak value) represent the thermal emission reconstructed in the 5-9 keV range. For both energy intervals, the data were integrated over 48 s during the onset of the impulsive phase. In the left panel, the blue crosses represent the error bars on the $x$ and $y$ coordinates of each source location provided by forward fit. We indicate the four sources from left to right with Fp 1, Fp 2, Fp 3, and Fp 4. A total number of 584,271 and 11,189 counts were measured by the imaging detectors in the selected thermal and nonthermal energy ranges, respectively. The yellow box in the middle panel indicates the field of view shown in Fig. \ref{Fig: thermal geometric comparison}.}
   \label{Fig: non-thermal STIX image}%
    \end{figure*}

   \begin{figure*}[htbp!]
   \centering
   \includegraphics[trim=5 100 5 100, clip,width=18cm]{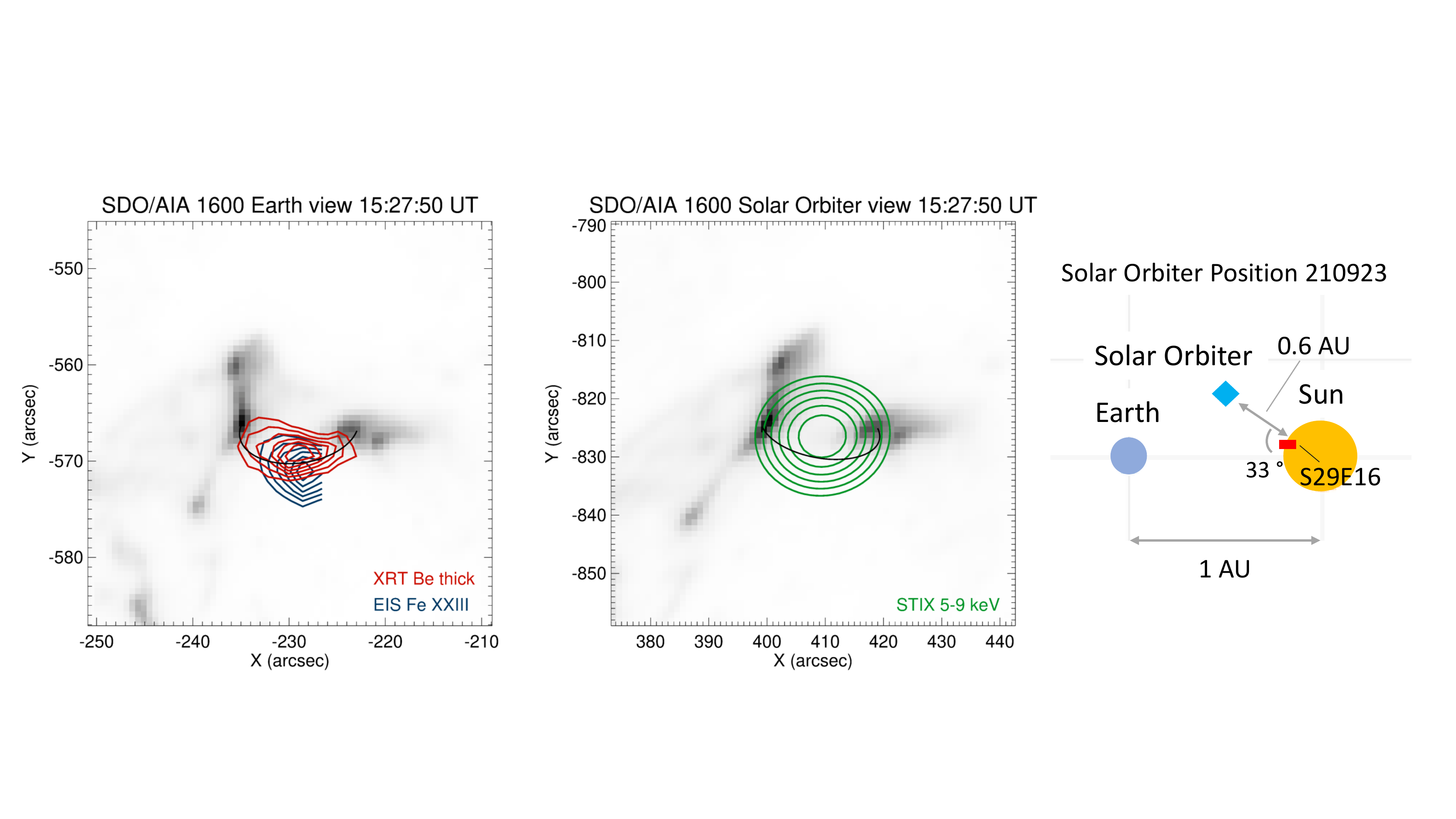}
   \caption{Comparison between the Earth and the Solar Orbiter vantage point during the 23 September 2021 flaring event (\emph{left} and \emph{middle} panel, respectively). In the left panel, the 40-90\% contours of the XRT Be thick filter (red) and the contours of the emission line \ion{Fe}{xxiii} of EIS (blue) are overlaid on the AIA 1600 \r{A}. In the middle panel, the 40-90\% contours of the STIX image of the thermal emission (green) reconstructed in the 5-9 keV range are overlaid on the reprojected AIA map. The black loop in the left panel is drawn to indicate a potential flare loop. The same black loop is shown in the middle panel, but reprojected according to the viewing angle of Solar Orbiter. These images were taken at the peak of the thermal emission of this flare, roughly 3 min later than the images shown in Fig. \ref{Fig: non-thermal STIX image}. The field of view of the middle panel is shown by the yellow box in Fig. \ref{Fig: non-thermal STIX image}. The right panel shows the relative locations of the Sun, Earth, and Solar Orbiter. The distance of Solar Orbiter to the Sun is 0.6 AU.}
   \label{Fig: thermal geometric comparison}%
    \end{figure*}

\section{Observations} \label{Ch: Observations}

\subsection{Four nonthermal sources observed with STIX}
    
    The detailed study of the M1.8 GOES class flare SOL2021-09-23UT15:28 was motivated by the image reconstructed from STIX data in the nonthermal energy range, shown from 22 to 28 keV. In Fig. \ref{Fig: non-thermal STIX image} the contours for the three reconstruction methods we used are shown. Four distinct sources are clearly visible in the AIA 1600 \r{A} filter and in the reconstructed images of STIX. This raises the question how these four sources can be brought into the context of the standard model of solar flares. The two inner sources are connected by thermal emission seen in 5-9 keV (red contours in Fig. \ref{Fig: non-thermal STIX image}), indicating that these two footpoints are likely to be the main flare ribbons. The further analysis was carried out to understand the behavior of the four sources and determine differences and/or similarities between them. 

   \begin{figure*}[htbp!]
   \centering
   \includegraphics[trim=5 60 5 50, clip,width=18cm]{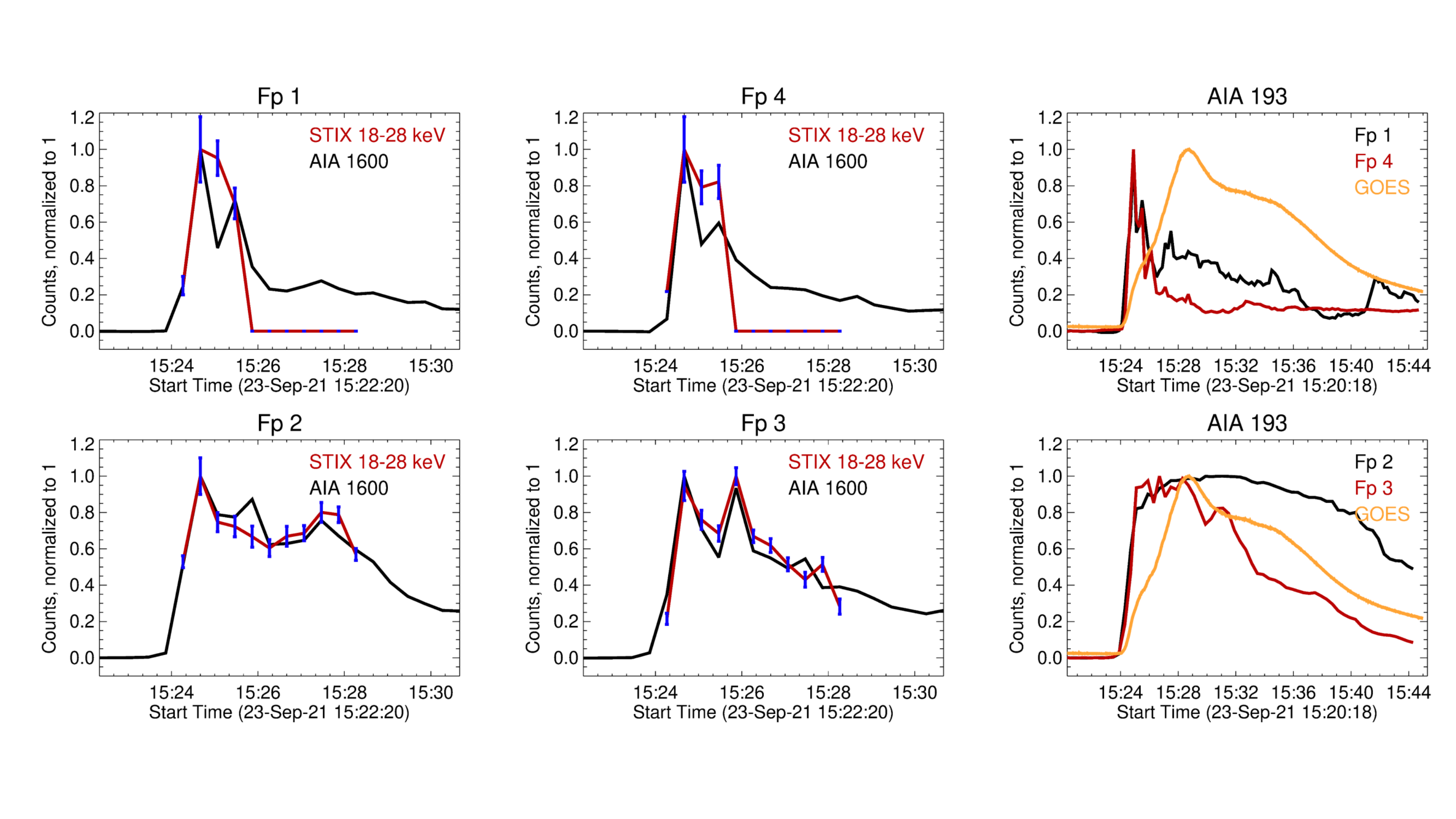}
   \caption{Time evolution plots with maximum value normalised to one. The left and middle column are the intensity plots of AIA 1600 \r{A} (black) and nonthermal STIX (red, with error bars in blue) for the four sources individually. The right column shows the intensity plots of AIA 193 \r{A} (black and red) for the four sources. As a reference, we added the GOES curve (orange).}
   \label{Fig: Time evolution}%
    \end{figure*}

\subsection{Geometric comparison between Solar Orbiter/STIX, SDO/AIA, Hinode/XRT, and Hinode/EIS} \label{Ch: geometry}
    
    In a first step, we compared the thermal sources observed by the instruments XRT, EIS, AIA, and STIX at the peak time of the thermal emission to understand the geometry of the flare. The left panel of Fig. \ref{Fig: thermal geometric comparison} shows the UV emission measured by AIA 1600 \r{A}. The contours in the left panel represent the hot flare loop as seen by XRT and EIS. We used the XRT Be thick filter because the Be med and the Be thin filters where saturated during the flare. For EIS, we show the \ion{Fe}{xxiii} emission line because \ion{Fe}{xxiv} was partially saturated, but it is clear despite the saturation that the emission of \ion{Fe}{xxiv} comes from the same location as \ion{Fe}{xxiii}. The middle panel shows the same UV emission, but the image is reprojected to the viewing angle of Solar Orbiter. The contours give the intensity measured by STIX in the thermal energy range of 5-9 keV using MEM\_GE for the reconstruction.
    
    The emission measured by XRT in the soft X-ray range shows a coronal loop that connects the two inner sources seen in UV (Fp 2 and Fp 3 in Fig. \ref{Fig: non-thermal STIX image}). To illustrate a potential flare loop, we added the loop in black that connects the UV footpoints and traces the soft X-ray loop. The black loop in the middle image represents the same loop as in the left image, but reprojected to the Solar Orbiter viewing angle. This helps to understand the 3D geometry of the flare and connects the two different viewing angles. The emissions of \ion{Fe}{xxiii} and of the thermal STIX reconstruction (5-9 keV) both come from the coronal loop top. The centroid of the thermal hard X-ray loop is offset by a few arcseconds relative to the black reference loop. This offset is within the current accuracy of the STIX aspect solution, and therefore it is not possible to draw a conclusion. Altogether, this analysis shows us that Fp 2 and Fp 3 in Fig. \ref{Fig: non-thermal STIX image} are the footpoints of a flare loop, in accordance with the standard model for solar flares.

\subsection{Intensity time evolution of the sources}

    To determine the differences and similarities of the four sources, we investigated their time evolution by measuring the intensity of the flux emitted at the locations of the four sources in different wavelengths as a function of time. To do this, we used the measurements of AIA 1600 \r{A}, AIA 193 \r{A}, and nonthermal STIX (18-28 keV). For the two AIA filters, we defined boxes around the four sources and extracted the intensity for each frame. Furthermore, we normalized the intensity by the corresponding exposure time. For AIA 1600 \r{A}, we obtained a 24 s cadence and for AIA 193 \r{A} a 12 s cadence. Because every second frame (24 s cadence) of AIA 193 \r{A} was strongly saturated due to the automatic exposure time adjustment of AIA, however, we also used a 24 s cadence for this filter as we only considered nonsaturated images. For STIX we reconstructed an image for each time point when there was an AIA 1600 \r{A} frame using forward fitting. As we know from the initial forward fitting, see Fig. \ref{Fig: non-thermal STIX image}, where the four sources are located, we fixed the source locations and sizes, and we only fit their intensities. This was done to increase the accuracy of the retrieved fluxes and to decrease the associated errors. Using the output fluxes and errors of the forward-fitting method, we constructed time profiles of the fluxes for the individual sources. They are plotted in Fig. \ref{Fig: Time evolution}.
    
   \begin{figure*}[htbp!]
   \centering
   \includegraphics[trim=150 80 140 105, clip,width=17cm]{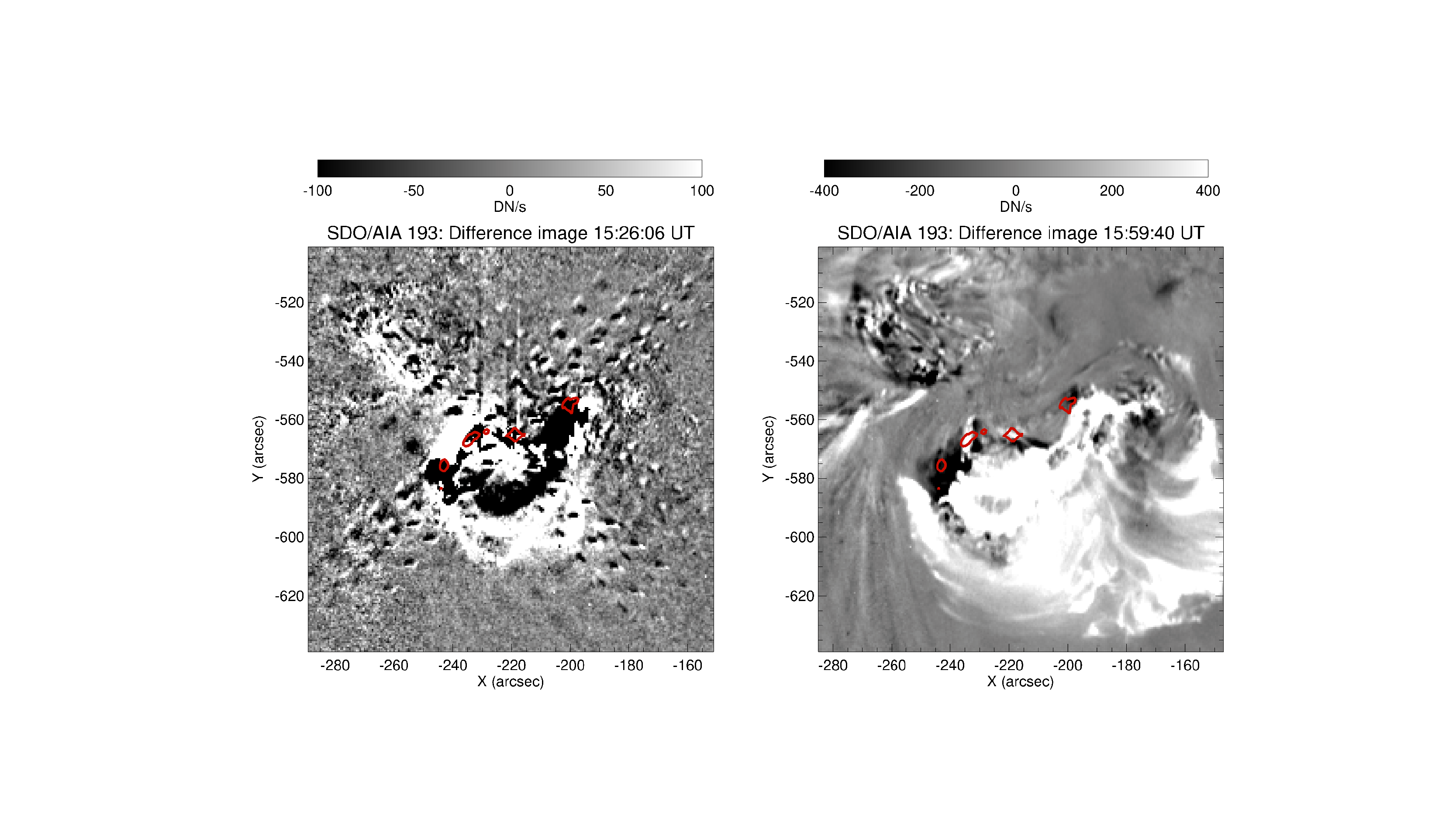}
   \caption{Difference images of the AIA 193 \r{A} filter at two different times to visualize the propagation of the CME. The left panel is an image from a running-difference movie during the impulsive phase of the flare, and the right panel is a difference image taken 35 minutes after the flare onset. A preflare image is subtracted to identify the dimming regions. The diagonal lines in the left panel are diffraction patterns. The red 40\% contours show the location of the four sources as seen by AIA 1600 \r{A} in the impulsive phase for reference.}
   \label{Fig: CME location}%
    \end{figure*}
    
    The plots show that the two inner sources (Fp 2 and Fp 3) behave similarly. Initially, the intensity rises fast, and then there is a longer cooling time. As each filter and instrument observes plasma at different temperatures, we measure different time profiles and especially different cooling curves, depending on which filter we are looking at. For example, for AIA 193 \r{A}, we see a long cooling time. This is the behavior we expect for footpoints of a coronal loop, and it supports our observations from Sect. \ref{Ch: geometry}. The inner sources are significantly different from the outer sources (Fp 1 and Fp 4). The outer sources are only visible at the onset of the flare and rapidly disappear afterward in a time range of roughly 2-3 min. This can be seen in the nonthermal images of STIX, in AIA 1600 \r{A}, and in AIA 193 \r{A}. This indicates that the outer sources represent a physically different process than the inner sources, which are the footpoints of a coronal loop. \citet{Chen_2020} observed anchor points of an erupting flux rope in the MW range. They described a similar behavior in the time evolution of the anchor points. The source regions are visible at the onset of the flare, followed by a rapid decay.
    
\subsection{Study of the CME in the low corona using AIA} 
\label{Ch: Study of CME}

    As there was a CME starting around 15:25 UT in the flaring region, we compared the location of the flare and the four footpoint-like sources with the CME and its propagation. The motivation for this was to investigate whether the four sources were connected to the CME. For the following analysis, we used the AIA 193 \r{A} filter. We created running-difference movies to understand the propagation of the CME, and we subtracted preflare images from images after the onset of the flare to identify dimming regions \citep{Harrison_2000}. As above for the time evolution, we only used every second, nonsaturated image of AIA 193 \r{A}, leaving us with a cadence of 24 s. In Fig. \ref{Fig: CME location} two image examples are shown. In the left panel, an image of the running-difference movie is shown, and in the right panel, an image for the dimming regions is shown.
    
    The left image in Fig. \ref{Fig: CME location} shows the outward movement of the CME. The loop-like structure that is moving outward seems to be bounded by the two outer sources (Fp 1 and Fp 4). The crossed pattern is a diffraction artifact. The right image shows that the outer sources lie in the dimming regions. At this time, it is particularly evident for Fp 1. When these two images are combined, it seems as if the outer sources are connected to the CME and material was moved away from the positions of the outer sources. 
    
    As an eruption of a filament might have caused the CME, we searched for a filament using the AIA 171 \r{A} and AIA 304 \r{A} filters. We were not able to observe an active filament at the flare location that might have been involved in the flare. Hot plasma is likely lying above the filament, which makes it difficult to observe the filament with its cold plasma. 

   \begin{figure*}[htbp!]
   \centering
   \includegraphics[trim=0 0 0 0, clip,width=18cm]{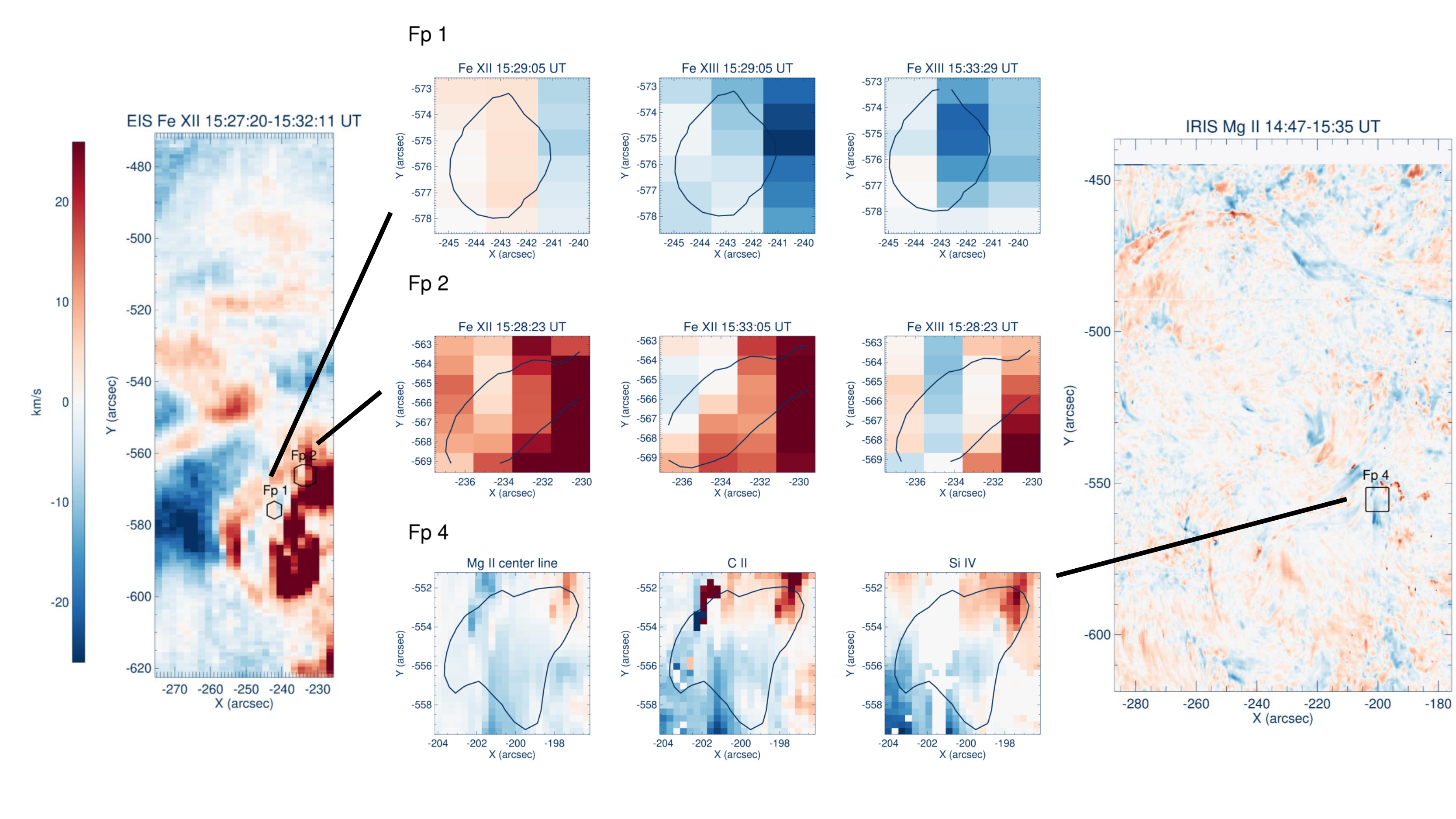}
   \caption{Doppler velocities measured by EIS and IRIS. The leftmost and rightmost panels are the whole raster frames from EIS and IRIS, respectively. In the frames we mark the regions around Fp 1, Fp 2, and Fp 4. These regions are shown in more detail for different emission lines and at different times in the nine middle panels. The top row of the compilation shows the measurement of EIS around Fp 1, the middle row the measurement of EIS around Fp 2, and the bottom row the measurement of IRIS around Fp 4. The dark blue contours are the 20 \% contours of AIA 1600 \r{A} showing the three sources Fp 1, Fp 2, and Fp 4.}
   \label{Fig: Vel plots}%
    \end{figure*}

\subsection{Measurement of the Doppler velocities at the source locations} \label{Ch: Velocities}
    
    Finally, we analyzed the Doppler velocities measured at the source positions. For Fp 1 and 2, we have EIS coverage at two time points, and for Fp 4 we have IRIS coverage. For Fp 3 we lack observational coverage entirely. In Fig. \ref{Fig: Time evolution for Vel} we mark the times for which we have measurements of the velocities for the individual sources. Taking the velocity maps of EIS and IRIS, we selected the regions around the three sources. In Fig. \ref{Fig: Vel plots} we show a compilation of the velocity maps at the three footpoints mentioned above. In Table \ref{Tab: Velocity values} we list the corresponding mean velocity values for these nine images.
    
    The outer sources (Fp 1 and Fp 4) are blueshifted ,whereas the inner source Fp 2 is redshifted, although the values are rather low, as Table \ref{Tab: Velocity values} shows. This supports the observations we described so far. As Fp 2 is the footpoint of a flare loop, the redshifted measurement at this location for the emission lines \ion{Fe}{xii} and \ion{Fe}{xiii} is to be expected (e.g., \citet{Milligan_2009}). The measured blueshift at the outer sources supports the  observation we described in Sect. \ref{Ch: Study of CME}. Plasma seems to be moving upward from the source locations \citep{Harra_2007}. The blueshift in the filters of IRIS indicates that cold plasma is moving. This suggests an involvement of a filament eruption that is connected to the outer sources Fp 1 and Fp 4. The nonthermal velocities measured by EIS at Fp 1 and Fp 2 are at the same level and show no differences \citep{Harra_2013}.
    
   \begin{table}
      \caption[]{Mean velocity values of the nine panels shown in the middle of Fig. \ref{Fig: Vel plots}. The order corresponds exactly to the order of the images in Fig. \ref{Fig: Vel plots}.}
         \label{Tab: Velocity values}
         $$
         \begin{array}{c|c|c|c}
            \hline\hline
            \noalign{\smallskip}
            \mathrm{Velocities}  & \mathrm{Left} & \mathrm{Middle}  & \mathrm{Right}\\
            \mathrm{[km/s]} & & &\\
            \noalign{\smallskip}
            \hline
            \noalign{\smallskip}
            \mathrm{Fp\;1} & -0.3 \pm 0.3 & -10.4 \pm 0.8 & -7.9 \pm 0.9\\
            \mathrm{Fp\;2} & 18.1 \pm 0.2 & 12.4 \pm 0.1 & 4.6 \pm 0.4\\
            \mathrm{Fp\;4} & -2.6 & -1.9 & -0.6\\
            \noalign{\smallskip}
            \hline
         \end{array}
     $$ 
   \end{table}
%

   \begin{figure*}[htbp!]
   \centering
   \includegraphics[width=15cm]{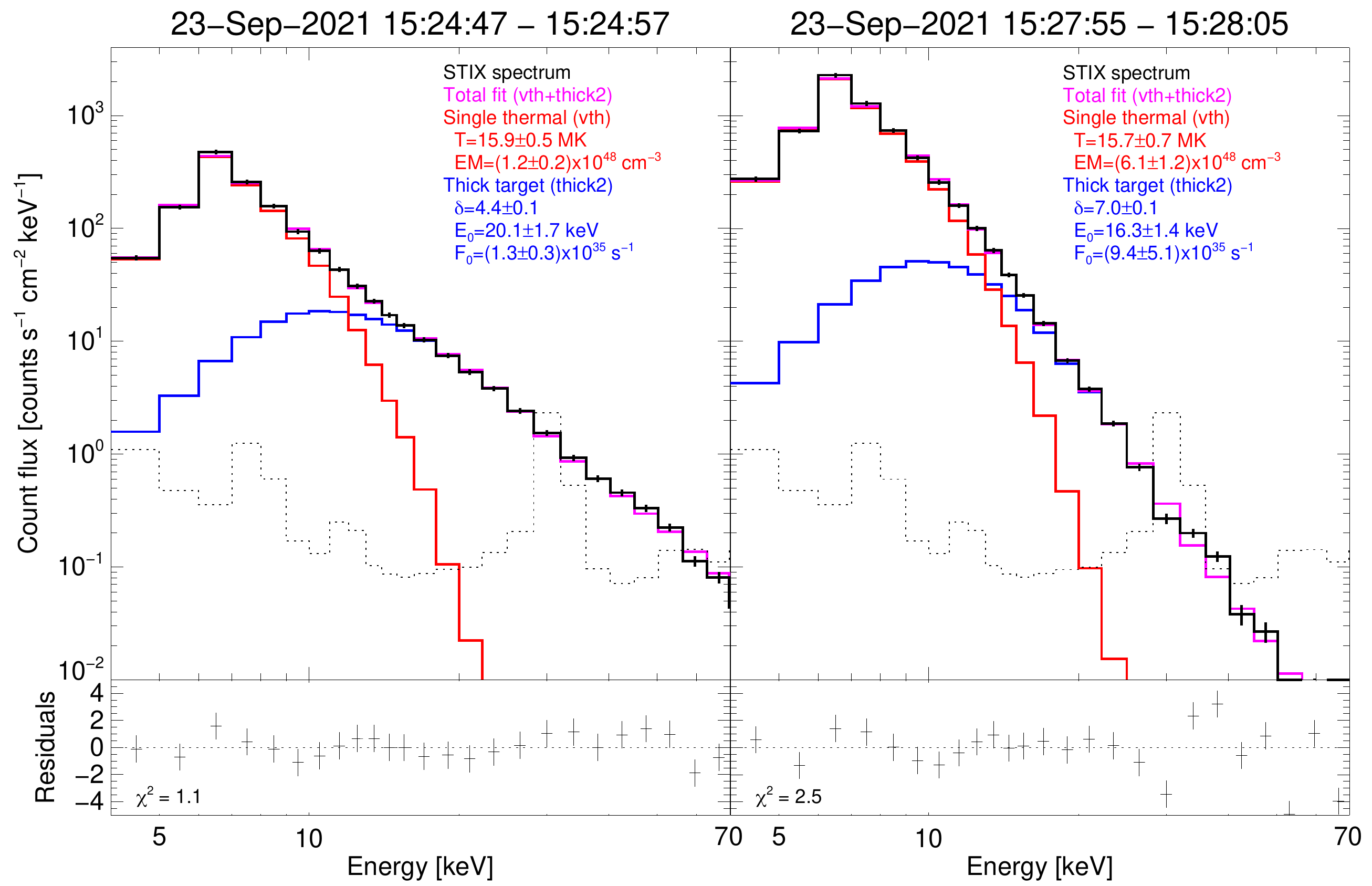}
   \caption{STIX background-subtracted count flux spectra around the nonthermal (\emph{left}) and thermal (\emph{right}) peaks. Both spectra can be well fit with an isothermal (red) component plus a thick target model (blue). The magenta curves correspond to the sum of the two components, and the dashed black curves represent the STIX background spectra taken during nonflaring times close to the event. Below each spectrum, we report the residuals (observations minus total fit) in units of the standard deviation calculated from counting statistics. The resulting fit parameters are shown in the legend of each plot.}
   \label{Fig: observed STIX spectrum}%
    \end{figure*}   
    
\subsection{STIX imaging spectroscopy of the nonthermal sources}
    The observed STIX X-ray spectrum can be well fit by a isothermal component and a thick target model using standard spectroscopy techniques (e.g., \citet{Battaglia_2021}). Figure \ref{Fig: observed STIX spectrum} shows the observed STIX spectrum in black with the thermal and nonthermal fit in red and blue. From the thick target model fit, we can read out the power-law index of the electron spectra $F\propto E^{-\delta}$. For the nonthermal peak, the power-law index is given by $\delta = 4.4\pm 0.1$. From this, we can determine the power-law index $\gamma$ of the photon spectra ($F\propto E^{-\gamma}$) given by $\gamma=\delta +1=5.4 \pm 0.1$.
    
   \begin{table}
      \caption[]{Results for the power-law index $\gamma$ for each footpoint individually. For the reconstruction, we used the forward-fitting method. We calculated $\gamma$ twice, first using the energy ranges [22,25] keV and [25,28] keV (\emph{middle column}), and then the energy ranges [22,25] keV and [25,32] keV (\emph{right column}). }
         \label{Tab: STIX imaging spectroscopy}
         $$
         \begin{array}{l|ll}
            \hline\hline
            \noalign{\smallskip}
            \mathrm{Energy\; range}  & 22-28 \;\mathrm{keV} & 22-32 \;\mathrm{keV} \\
            \noalign{\smallskip}
            \hline
            \noalign{\smallskip}
            \mathrm{Fp\;1} & 5.1 \pm 2.6 & 7.1 \pm 1.5\\
            \mathrm{Fp\;2} & 6.8 \pm 1.4 & 4.6  \pm 0.9\\
            \mathrm{Fp\;3} & 8.0 \pm 1.3 & 6.0 \pm 0.8\\
            \mathrm{Fp\;4} & 7.0 \pm 2.3 & 4.6 \pm 1.4\\
            \noalign{\smallskip}
            \hline
         \end{array}
     $$ 
   \end{table}

    In the following, we discuss our efforts to derive the spectra for the individual sources to determine whether there are differences or similarities between the spectra of the inner and outer sources. To do this, we needed to separate the nonthermal energy range into at least two intervals, and we separately derived images for each range. This was a challenge because we had to reconstruct four sources with about half the counts compared to the images shown in Fig. 3. To create images, we used the forward-fitting method because it provides clearly defined error estimates on the flux of each source. To facilitate the fitting, we fixed the source locations and sizes and only fit the fluxes. The source locations were taken from the forward-fit result over the entire nonthermal energy range, while the source size was fixed at the nominal resolution of the finest subcollimator we used ($\sim$15$''$). This approach is justifiable because it is only useful to derive individual source spectra if the sources at all energies are from the same location. 
    
    To determine the power-law index, we used two energy ranges, with $E_1$ and $E_2$ being the mean values of these energy ranges. As it is not clear a priori  how to break up the nonthermal energy range, we calculated $\gamma$ for two different sets of energy ranges, the energy ranges [22,25] keV and [25,28] keV, and the energy ranges [22,25] keV and [25,32] keV. The reconstructions performed with the forward fit in each energy range returned the corresponding fluxes $F_1$ and $F_2$ [s$^{-1}$keV$^{-1}$]. From $F_1$, $F_2$, $E_1$, and $E_2$, we were able to calculate the power-law index $\gamma$ using the equation
    \begin{equation}
        \label{Eq: Energy spectrum}
        \gamma=-\frac{\log{F_1}-\log{F_2}}{\log{E_1}-\log{E_2}}.
    \end{equation}
    
    Initially, we used the mean value of the energy ranges $E_1$ and $E_2$ to calculate $\gamma$ (i.e., assuming a flat spectrum) using Eq. \ref{Eq: Energy spectrum}. As we know that the spectrum is not flat, we used in a second step the initially calculated value of $\gamma$ to obtain values of $E_1$ and $E_2$ that were more accurate. We used the weighted average over the spectral shape for this. Then $\gamma$ was recalculated using Eq. \ref{Eq: Energy spectrum} together with the weighted values $E_1$ and $E_2$. In Table \ref{Tab: STIX imaging spectroscopy} the results for $\gamma$ are given for each source individually. The error estimation of $\gamma$ was made using the error of the flux returned by the forward-fitting method and error propagation. 
    
     Table \ref{Tab: STIX imaging spectroscopy} shows that the errors are very large, which results in a rather high spread for the power-law index $\gamma$. This highlights that a STIX image reconstruction of four sources is challenging for these low counts (i.e., about 5000 counts per image). The obtained values agree roughly with the power-law index we derived from the thick target model ($\gamma=5.4\pm 0.1$). The large errors prevent us from comparing the spectra of the inner and outer sources.

\section{Discussion and conclusions} \label{Ch: Discussion and conclusion}
    
\begin{figure}
   \centering
   \includegraphics[trim=260 20 160 0, clip,width=9cm]{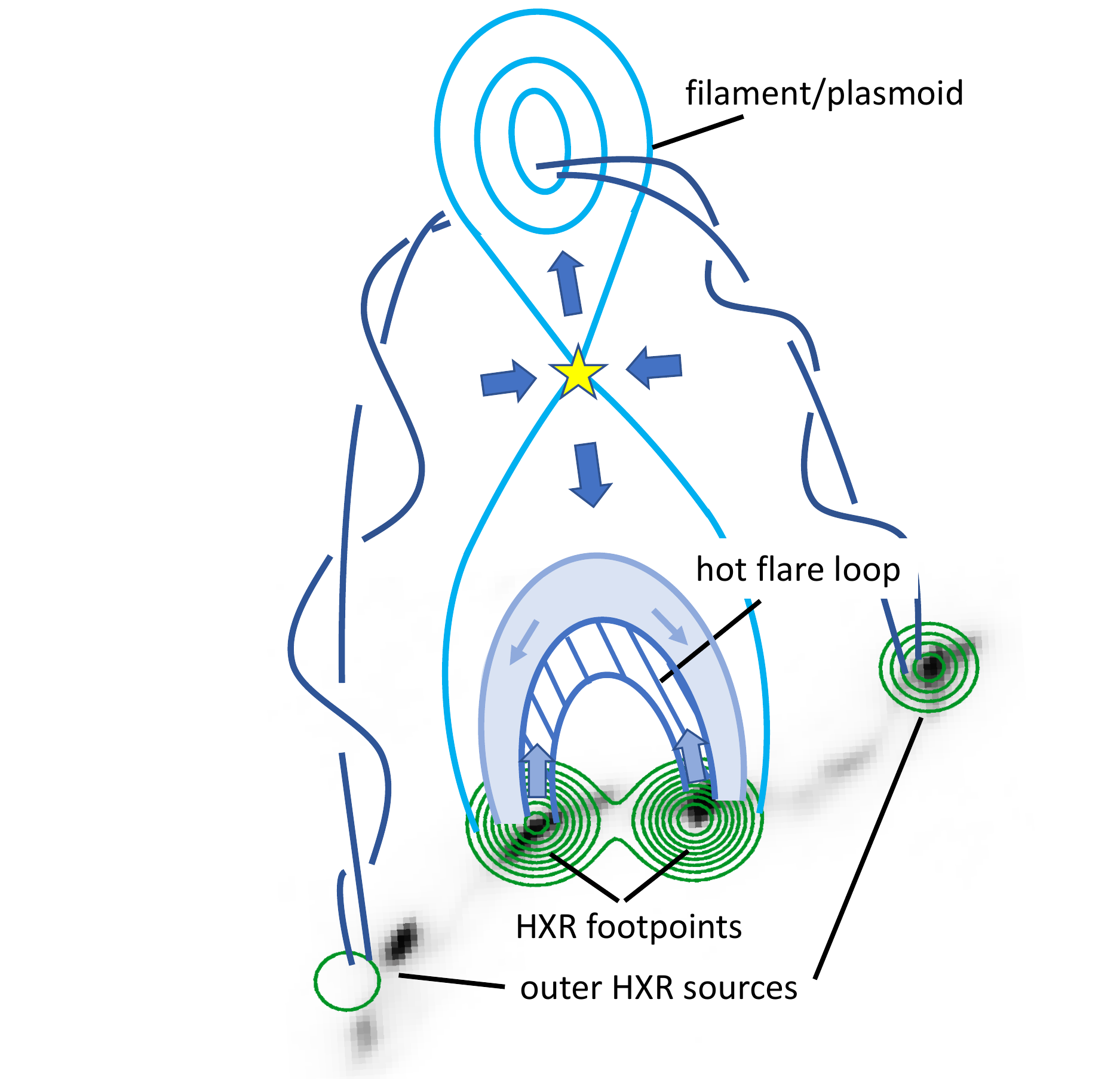}
      \caption{Cartoon of the presented flare during the onset of the impulsive phase. The background image, which is the same as the left panel of Fig. \ref{Fig: non-thermal STIX image}, shows the contour levels of the four sources observed by STIX in the nonthermal energy range 22-28 keV, overlaid on the reprojected AIA 1600 \r{A} map. Additionally, the standard model for solar flares is represented. The outer sources in hard X-rays from the anchor points of the erupting filament are the new contribution to the standard flare picture.}
    \label{Fig: Summary with standard model}
   \end{figure}

    The analysis presented in this paper was motivated by the four sources that we reconstructed in the nonthermal energy range using STIX data. The goal was to understand the behavior of these four sources and bring them into the context of the standard model of solar flares. To do this, we investigated the flare geometry and compared the time evolution and velocities at the four sources. Overall, we observed that we can split up the four sources into the inner sources, Fp 2 and Fp 3, and the outer sources, Fp 1 and Fp 4, based on their characteristics and behavior. The observations for the inner sources revealed that they are the footpoints of flare loops, as is consistent with the standard model. We concluded this from their location in the flare when compared to the location of the thermal sources observed at other wavelengths, from their time evolution, and from the velocity measurement at Fp 2. The outer sources behave very differently than the inner sources in a few respects. We observed a fast increase and decrease in the intensity time profiles, showing that the outer sources exist for a shorter time than the inner sources. A comparison of the locations of the outer sources with the movement of the contemporaneous CME strongly indicates a link between the CME and the outer sources. This is supported by the blueshift measurements at the locations of the outer sources. However, we were not able to identify the erupting filament in any of the AIA images. Combining everything, we deduce that the outer sources are likely the anchor points of an eruptive filament. The outer sources are produced by flare-accelerated electrons that precipitate down the filament and emit Bremsstrahlung in the chromosphere. Electrons streaming down the filament also radiate in microwaves through the gyrosynchrotron process, as has been reported by \citet{Chen_2020} for SOL2017-09-10. No microwave observations are available for the flare analyzed in this paper. Similarly as reported by \citet{Chen_2020} for MW, the outer HXR sources last shorter than the main emission from the footpoints of the flare loop. \citet{Chen_2020} suggested that this is because the access of flare-accelerated electrons to the chromosphere becomes increasingly difficult as the filaments move upward. In Fig. \ref{Fig: Summary with standard model} we summarised our result, showing the standard model of solar flares combined with the measurement of the flare on 23 September 2021. 
    
    Imaging four individual sources with STIX is pushing the limits of the STIX imaging capabilities giving the very limited number of visibilities measured by the instrument. However, for this flare, the similar intensities of the different sources made it possible to image them. A comparison of the imaging results provided by different methods, reported in Appendix \ref{Appendix A}, shows that the reconstructions are consistent in the different methods, thus demonstrating their reliability. A total of 11'189 counts (on top of a background of 565 counts) over the energy range 22 to 28 keV were used to reconstruct the images shown in Fig. \ref{Fig: non-thermal STIX image}. In Sect. 3.6 we further broke up the energy range and tried to use the reconstructed images to derive the individual spectra of the four sources. In these smaller energy ranges, we did not have sufficient counts to be able to reconstruct reliable fluxes of all four sources. As a rule of thumb, imaging reconstruction of four sources with similar intensities requires about 10'000 counts or more. For simpler images such as two footpoints, a few thousand counts are generally enough to reconstruct an image. 
    
    An interesting question to ask is why we never observed a flare like this within the extensive RHESSI data set, the predecessor of STIX. The imaging dynamic range and the sensitivity of RHESSI and STIX are roughly similar, at least for STIX observations made significantly closer to the Sun than 1 AU. If there were RHESSI observations of the flare presented in this paper, the four sources would also have been individually imaged by RHESSI. The absence of any published RHESSI flares with emissions from the anchor points of the filament suggests that the outer sources might be generally weaker than reported here. For outer sources that have intensities lower than 10\% of the peak, both RHESSI and STIX are unlikely to detect all four sources. In any case, to find the outer sources required a dedicated search, and a detailed analysis of the significance of the reconstructed sources needs to be performed, such as in the appendix of this paper. We are not aware of a systematic search of the RHESSI flare list for outer sources associated with an erupting filament, but the extensive RHESSI flare list with $\sim$600 M class flares is likely to include similar events as reported here.
    
    Based on the discussion above, we suggest that we might systematically search for similar flares in which we potentially see emission from the anchor points of a filament. Using the STIX data alone, we should concentrate our search on larger events with more than 10'000 counts in the nonthermal range. Then we need to constrain the time and energy range and investigate the significance of the different reconstructed sources (see the appendix). As the outer sources are visible in the AIA UV and EUV filters, especially the AIA 1600 \r{A} filter, the search could also start by scanning AIA flare data and look for an erupting filament with clearly identified anchor points. Scanning through the AIA images might be faster and easier. The AIA search will provide a selection of candidate flares that might be searched for outer X-ray sources. For the case of a nondetection, this method using AIA images has the advantage that an upper limit of the X-ray intensities can be derived. 
    
    In conclusion, this paper presented an analysis of an M-GOES class flare that was observed by IRIS, by SDO with AIA, by Hinode with EIS and XRT, and by Solar Orbiter with STIX. The observed eruptive flare shows the classical picture of a flaring loop and footpoints. Additionally, we measured emission away from the main flare loop in UV, EUV, and X-ray, showing nonthermal emission. These two outer sources show a similar spatial and temporal behavior that is different to the behavior of the two footpoints of the flare loop. We suspect that these outer sources, visible in the hard X-ray range, might be the anchor points of an erupting filament.

\begin{acknowledgements}
      \em{Solar Orbiter} is a space mission of international collaboration between ESA and NASA, operated by ESA. The STIX instrument is an international collaboration between Switzerland, Poland, France, Czech Republic, Germany, Austria, Ireland, and Italy. AFB, HC, and SK are supported by the Swiss National Science Foundation Grant 200021L\_189180 and the grant 'Activités Nationales Complémentaires dans le domaine spatial' REF-1131-61001 for STIX. PM acknowledges the financial contribution from the agreement ASI-INAF n.2018-16-HH.0 and from NASA Kentucky under NASA award number 80NSSC21M0362. Hinode is a Japanese mission developed and launched by ISAS/JAXA, with NAOJ as domestic partner and NASA and UKSA as international partners. It is operated by these agencies in co-operation with ESA and NSC (Norway). SDO data are courtesy of NASA/SDO and the AIA, EVE, and HMI science teams. IRIS is a NASA small explorer mission developed and operated by LMSAL with mission operations executed at NASA Ames Research center and major contributions to downlink communications funded by ESA and the Norwegian Space Centre.
\end{acknowledgements}

\bibliographystyle{aa} 
\bibliography{biblio} 

\begin{appendix}
\section{Robustness check of the reconstructed image with four sources } \label{Appendix A}

   \begin{figure*}[htbp!]
   \centering
   \includegraphics[trim=120 150 120 150, clip,width=18cm]{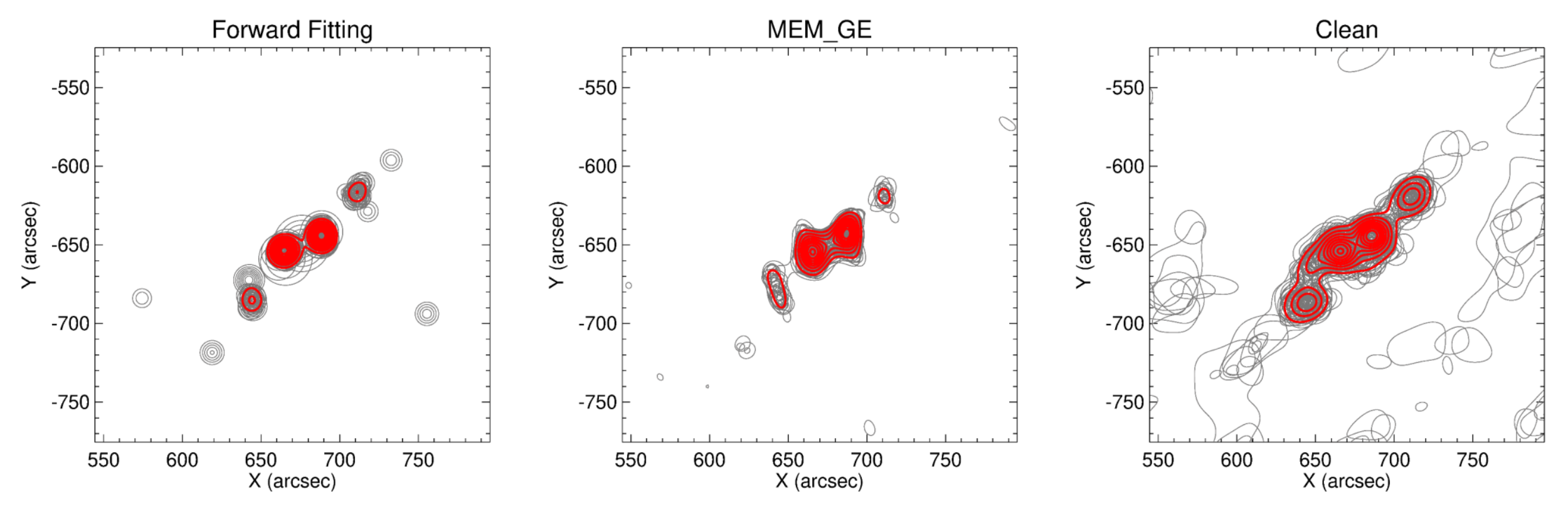}
   \caption{Results of the 20 reconstructions from perturbed data to determine the robustness of the imaging methods for STIX. The gray contours show the 20 reruns, and the red contours are the average over all runs. The contour levels are 20-90\% of the maximum values. For the images, we used the time range 15:24:28 UT to 15:25:16 UT (these times are corrected for the light travel time to Earth, about 198.8s) and the energy range 22-28 keV.}
   \label{Fig: error test Mean}%
    \end{figure*}
    
    Although the STIX data calibration has reached a good level of accuracy \citep{Massa_2022}, efforts to test and improve the current calibration are still ongoing. The flare discussed in this paper is the first event for which STIX detected four individual sources, and it provides an excellent test for the STIX imaging capabilities. To test the robustness of the reconstructions of the four individual sources, we applied the confidence strip method \citep{pianabook}. Twenty times, Gaussian noise with a standard deviation set equal to the error on the visibility amplitudes was added to the observed visibilities. Then, 20 reconstructions were performed using the perturbed data. The reconstructions were made using the same parameters as for the images in Fig. \ref{Fig: non-thermal STIX image}. In Fig. \ref{Fig: error test Mean} the 20 reconstructions are shown for the three reconstruction methods: forward fit, MEM\_GE, and Clean. In general, the image reconstruction of four sources is robust. Out of 20 reruns, the forward-fitting method three times failed to find the location of Fp 4 and to find the location of Fp 1 once, but otherwise, it gave consistent results. As a further test, we fixed the source sizes in the forward fit to 15 arcsec and repeated the 20 runs. With this restriction, there was no failure in the 20 runs. For the MEM\_GE and Clean methods, the four sources can be observed in all 20 runs. When we created the Clean images, the residuals were added to the final image at the end of the reconstruction process, while for the other methods, the residuals were not added. The addition of the residuals is the reason for the higher noise in Clean images (the right panel of Fig. \ref{Fig: error test Mean}). In summary,  we can reconstruct four sources, but as the forward fit fails to reconstruct the fourth sources in 3 out of 20 cases and the first source in one out of 20 cases, this is likely the limit of feasibility for STIX, at least for the number of counts that were available for this flare (roughly 10'000).
\end{appendix}

\end{document}